\documentstyle[twocolumn,floats,aps,epsf]{revtex}
\newcommand{\ab}{{\bf a}}
\newcommand{\bb}{{\bf b}}
\newcommand{\cb}{{\bf c}}
\newcommand{\bnu}{{\bf 0}}
\newcommand{\rb}{{\bf r}}
\newcommand{\eb}{{\bf e}}

\newcommand{\Ri}{{{\bf R}_i}}
\newcommand{\Ab}{{\bf A}}
\newcommand{\Bb}{{\bf B}}
\newcommand{\Cb}{{\bf C}}
\newcommand{\Ec}{{\cal E}}
\newcommand{\mPhi}{{\mit\Phi}}
\newcommand{\mOmega}{{\mit\Omega}}
\newcommand{\mGamma}{{\mit\Gamma}}
\newcommand{\sump}{\mathop{{\sum}'\!\!\!}}
\begin{document}
\twocolumn[\hsize\textwidth\columnwidth\hsize\csname@twocolumnfalse%
\endcsname
\title{Crystal potentials under invariant periodic boundary conditions at
infinity}
\author{Eugene V. Kholopov\cite{EVKhol}}
\address{Institute of Inorganic Chemistry of the Siberian Branch of
the Russian Academy of Sciences, 630090 Novosibirsk, Russia}
\maketitle
\draft
\begin{abstract}
The definiteness of bulk electrostatic potentials in solids under periodic
boundary conditions defined in an invariant manner has been proved in the
general case of triclinic symmetry. Some principal consequences following
from the universal potential correction arising are discussed briefly.
\end{abstract}
\pacs{PACS numbers: 41.20.Cv, 61.50.Lt}
]
\section{Introduction}
The summation of Coulomb potentials over crystal lattices is a
classical problem \cite{Glas80} important for determining the cohesive
energy in crystals \cite{Tosi64} as well as for describing electronic
properties there \cite{Harr75,Ihm988}. Despite a long history of this
subject ($\hspace{-3.5pt}$\cite{Khol01} and references therein) the
question about the uniqueness of electrostatic properties in the bulk
remains controversial. Indeed, it is well-known that every electrostatic
task is defied by boundary conditions \cite{Jack62} and periodic boundary
conditions are appropriate to the solution in the bulk \cite{Born12}. As a
result, any electric field generated by polar unit cells is to be excluded
as irrelevant to the bulk state \cite{Smit81,Deem90}. Notwithstanding, as
far as electrostatic potentials are concerned, an arbitrary constant
potential could be formally added as an available periodic solution of
Laplace's equation \cite{Harr75,Dahl65,Redl75,Zhan94}. Therefore, the
absence of a constant potential in the original approach of Ewald
\cite{Ewal21} is often treated as optional \cite{Harr75}, whereas the
wide-spread standpoint is that the electrostatic potentials are well
defined by boundary conditions on open surfaces of crystals, but these
potentials are undetermined to an additive constant in infinite crystals
\cite{Klei81}. This claim is, however, at variance with at least the two
physical statements: On the one hand, due to statistical arguments, the
bulk state cannot be governed by surface ones \cite{Born54,Grif68}.
Furthermore, the actual charge distribution in the bulk of a crystal is to
be in one-to-one correspondence with potentials it generates
\cite{Spac81,Gene87}.

In the present paper we show that periodic boundary conditions defined
properly are sufficient to make electrostatic potentials definite in the
bulk. It means that periodic boundary conditions are as effective as
boundary conditions on surfaces are. Moreover, the latter ones can then be
reconstructed as relevant to real surfaces consistent with bulk states
\cite{Tosi64,Task79,Wolf92}.

\section{Lattice summation as a convergent procedure}
In the general case of triclinic symmetry, let a crystal be described by
primitive translation vectors $\ab=a\eb_a$, $\bb=b\eb_b$ and $\cb=c\eb_c$,
where $\eb_a$, $\eb_b$ and $\eb_c$ are the appropriate unit vectors, with
the products $(\eb_a\eb_b)=\cos\alpha$, $(\eb_b\eb_c)=\cos\beta$ and
$(\eb_c\eb_a)=\cos\gamma$. In terms of a charge distribution $\rho(\rb)$
contained in a unit cell parallelepiped, the direct lattice-sum
contribution to the electrostatic potential at a reference point $\rb$ can
be written as
\begin{equation}\label{Aq1}
U_{\rm Cd}(\rb)=\sump_i\int_V\frac{\rho(\rb')\:d\rb'}{|\Ri+\rb'-\rb|} ,
\end{equation}
where $i$ runs over the corresponding Bravais lattice specified by $\Ri$,
the prime on the summation sign implies missing the singular contributions
of the summand, the integration is over the unit-cell volume.

To make the result of summation in (\ref{Aq1}) definite, the following
conditions for the absolute convergence of (\ref{Aq1}) are to be suggested
\cite{Evje32,Coog67}:
\begin{eqnarray}
&&{\displaystyle\int_V}\rho(\rb)\:d\rb=0 ,\label{Aq2}\\
&&M_\mu\equiv{\displaystyle\int_V}r_\mu\rho(\rb)\:d\rb=0 , \label{Aq3}\\
&&G_{\mu\nu}\equiv{\displaystyle\int_V}r_\mu r_\nu\rho(\rb)\:d\rb=0
\quad\mbox{at}\quad\mu\neq\nu ,\label{Aq4}\\
&&G_{xx}=G_{yy}=G_{zz}=H ,\label{Aq5}
\end{eqnarray}
where $r_\mu$ are Cartesian components of $\rb$, $H$ is an arbitrary
constant. If any initial $\rho^{\rm ini}(\rb)$ is not subjected to
(\ref{Aq3})--(\ref{Aq5}), provided that the neutrality condition
(\ref{Aq2}) holds, then we may modify $\rho^{\rm ini}(\rb)$ as follows:
\begin{equation}\label{Aq6}
\rho(\rb)=\rho^{\rm ini}(\rb)+\sum_jq_j\delta(\rb-\rb_j) ,
\end{equation}
where $\delta(\rb)$ is the Dirac delta function, $q_j$ are some fictitious
point charges \cite{Evje32,Fran50} located in the vicinity of the origin,
$\rb=0$, at points $\rb_j$ connected by admissible lattice translations.
The values of $q_j$ are constrained by
\begin{equation}\label{Aq7}
\sum_jq_j=0
\end{equation}
so as to exclude the contribution of $q_j$ to the overall initial charge
distribution upon combining $\rho(\rb)$ attributed to neighbouring unit
cells. It is significant that ten different charge species among $q_j$ are
sufficient to fulfil (\ref{Aq3})--(\ref{Aq5}) and (\ref{Aq7}). Thus, $H$
remains optional in (\ref{Aq5}).

It is advantageous to consider a unit cell with the origin in its
geometric centre. Then a compact distribution of $q_j$ is supplied by
$\rb_j$ belonging to the following set of vectors: $\bnu$, $\pm\ab$,
$\pm\bb$, $\pm\cb$, $\pm\ab\pm\bb$, $\pm\bb\pm\cb$ and $\pm\cb\pm\ab$.
Keeping in mind that $\rho^{\rm ini}(\rb)$ is still contained in the
unit-cell parallelepiped, $\rho(\rb)$ can be connected with a
parallelepiped that is the same in shape, but twice as large in size. It
is convenient to adopt this parallelepiped as a new unit cell. Being
additive, the potential of interest is then described by the sum of
independent contributions of eight interpenetrating lattices composed of
new unit cells each. Thus, it is conceptually sufficient to discuss the
potential effect generated by a single lattice specified by
(\ref{Aq1})--(\ref{Aq5}), provided that this lattice is still determined
by the lattice parameters $a$, $b$ and $c$, as is verified later on.

\section{Invariance of periodic boundary conditions}
To make the solution of interest determinate, we assume that the overall
structure is composed of an integral number of unit cells restricted by 
planes which are parallel to the unit-cell faces and are specified by the 
vectors $\pm\Ab$, $\pm\Bb$ and $\pm\Cb$ relative to a central unit cell, 
where $\Ab=A\eb_a$, $\Bb=B\eb_b$ and $\Cb=C\eb_c$ at
\begin{equation}\label{Aq8}
\frac{A}{a}=\frac{B}{b}=\frac{C}{c}\gg1 ,
\end{equation}
so that the uniformity along each crystallographic direction is
maintained. Periodic boundary conditions are then readily involved as
imposed in such a way that each couple of remote parallel restricting
planes merges, so that equal number of complete unit cells occur along
each direction of $\eb_a$, $\eb_b$ and $\eb_c$. In this event, the
invariant character of periodic boundary conditions implies that each
plane of merging may also occur somewhere within boundary unit cells,
without changing the result. It is important that if planes of merging
happen in intermediate positions specified by $0\leq f\leq1$ within
boundary unit cells, without loss of generality, some instantaneous charge
distributions are to be introduced on those planes so as to fulfil
conditions (\ref{Aq2}) and (\ref{Aq3}) furnishing the convergence of the
surface potential contributions. As a result, the contribution of any
dipolar polarization along those planes can be eliminated, but dipolar
moments normal to the planes in question are inevitable and contribute to
the potential value in the interior \cite{Task79,Lee980,Hey281,Coke83}. In
the particular case of the $+\Ab$ plane one can show that the
corresponding potential contribution takes the form
\begin{eqnarray}\label{Aq9}
\mPhi_{\Ab}(f)&=&\frac{\mOmega^2}{\sin\beta}\!\int_{-a/2}^{t(f)}\!
\!dt\!\!\int_{-b/2}^{b/2}\!\!dp\!\!\int_{-c/2}^{c/2}\!\!du\,
\rho(t,p,u)\nonumber\\
&&{}\times\Bigl[t(f)-t\Bigr]\sum\limits_{i\in\{\Ab\}}
\frac{R_i^{\bot\Ab}}{R_i^3} ,
\end{eqnarray}
where in the triclinic co-ordinates $\rb=t\eb_a+p\eb_b+u\eb_c$,
$t(f)=af-a/2$, $i$ runs over unit cells truncated by the boundary plane,
with $R_i^{\bot\Ab}$, the component of $\Ri$ along an outward normal to
this plane,
\begin{eqnarray}\label{Aq10}
\mOmega&=&\bigl[1-\cos^2\alpha-\cos^2\beta-\cos^2\gamma\nonumber\\
&&{}+2\cos\alpha\;\cos\beta\;\cos\gamma\bigr]^{1/2} .
\end{eqnarray}
Upon merging the $\pm\Ab$ planes, auxiliary charges on those planes cancel
each other at a given $f$, but the aforementioned invariance of the
boundary conditions is based on (\ref{Aq9}) averaged over $f$, so that the
effective potential generated by that couple of planes is to be specified
as
\begin{equation}\label{Aq11}
\bar{\mPhi}_{\Ab}=\int_0^1\Bigl[\mPhi_{\Ab}(f)+\mPhi_{-\Ab}(1-f)\Bigr]df ,
\end{equation}
where $\mPhi_{-\Ab}(f)$ follows from (\ref{Aq9}) upon inversion of the
co-ordinate system. Carrying out the integration over $f$ in (\ref{Aq11})
and utilizing (\ref{Aq5}), we obtain
\begin{equation}\label{Aq12}
\bar{\mPhi}_{\Ab}=\frac{H\sin\beta}{a\mOmega}\sum\limits_{i\in\{\Ab\}}
\frac{R_i^{\bot\Ab}}{R_i^3} .
\end{equation}

\section{Uniqueness of bulk potentials}
The structural factor described by the sum in (\ref{Aq12}) determines the
potential at a large distance from the plane at hand. Therefore, it is
independent of the discrete character of that sum \cite{Redl75,Task79,%
Coog67,Hey281} and so can be represented in the following integral form
\begin{equation}\label{Aq13}
\sum\limits_{i\in\{\Ab\}}\frac{R_i^{\bot\Ab}}{R_i^3}=\frac{a\mOmega}{bc
\sin\beta}\int_{-b}^b dp\int_{-c}^c\frac{du}{W^3(a,p,u)} ,
\end{equation}
where the scale transformation to the parameters of a unit cell is
performed in agreement with (\ref{Aq8}),
\begin{eqnarray}\label{Aq14}
W(a,b,c)&=&\bigl(a^2+b^2+c^2+2ab\cos\alpha+2bc\cos\beta\nonumber\\
&&{}+2ca\cos\gamma\bigr)^{1/2} .
\end{eqnarray}
Carrying out the integration in (\ref{Aq13}) and substituting the
result into (\ref{Aq12}), we derive
\begin{eqnarray}\label{Aq15}
\bar{\mPhi}_{\Ab}&=&\frac{H}{v}\Bigl[Y(a,\!b,\!c|\alpha,\!\beta,\!
\gamma)\!-\!Y(-a,\!b,\!c|\alpha,\!\beta,\!\gamma)\!\nonumber\\
&&{}-\!Y(a,\!-b,\!c|\alpha,\!\beta,\!\gamma)\!-\!
Y(a,\!b,\!-c|\alpha,\!\beta,\!\gamma)\Bigr] ,
\end{eqnarray}
where $v=abc\mOmega$ is the volume of the unit cell,
\begin{eqnarray}
&&Y(a,b,c|\alpha,\beta,\gamma)=\tan^{-1}\Bigl\{\bigl[bc\sin^2\!\beta
+ab\mGamma(\gamma)\nonumber\\
&&\qquad{}+ca\mGamma(\alpha)-a^2\mGamma(\beta)\bigr]\bigl[a\;
\mOmega\;W(a,b,c)\bigr]^{-1}\Bigr\} ,\label{Aq16}\\
&&\mGamma(\phi_1)=\cos\phi_1-\cos\phi_2\cos\phi_3 ,\label{Aq17}
\end{eqnarray}
the parameters $\phi_j$ are the angles $\alpha$, $\beta$ and $\gamma$ in
an arbitrary combination.

As a generalization of (\ref{Aq15}), the total potential contribution
associated with the $\pm\Ab$, $\pm\Bb$ and $\pm\Cb$ boundary planes after
their merging takes the form
\begin{equation}\label{Aq18}
\mPhi_{\rm top}=\bar{\mPhi}_\Ab+\bar{\mPhi}_\Bb+\bar{\mPhi}_\Cb ,
\end{equation}
where $\bar{\mPhi}_\Bb$ and $\bar{\mPhi}_\Cb$ are obtained from
(\ref{Aq15})--(\ref{Aq17}) upon the cyclic interchanges $\Ab\to\Bb\to\Cb$,
$a\to b\to c\to a$ and $\alpha\to\beta\to\gamma\to\alpha$ there. To
proceed further, we remark that the angles $Y(a,b,c|\alpha,\beta,\gamma)$,
$Y(b,c,a|\beta,\gamma,\alpha)$ and $Y(c,a,b|\gamma,\alpha,\beta)$
associated with $W(+a,+b,+c)$ and denoted as $\chi_j$ form a closed set
with the property
\begin{equation}\label{Aq19}
\tan(\chi_1+\chi_2)\tan\chi_3=1 ,\qquad |\chi_j|<\pi/2 .
\end{equation}
One can readily prove therefrom that
\begin{equation}\label{Aq20}
\chi^{+++}=\chi_1+\chi_2+\chi_3=\frac{\pi}{2} ,
\end{equation}
where the sign combination specifying the arguments of $W(+a,+b,+c)$ is
indicated as a superscript. Likewise, for the angles associated with
$W(-a,+b,+c)$, $W(+a,-b,+c)$ and $W(+a,+b,-c)$ we get
\begin{equation}\label{Aq21}
\chi^{-++}=\chi^{+-+}=\chi^{++-}=-\frac{\pi}{2} .
\end{equation}
Substituting (\ref{Aq20}) and (\ref{Aq21}) into (\ref{Aq18}), we finally
reach
\begin{equation}\label{Aq22}
\mPhi_{\rm top}=\frac{2\pi H}{v} .
\end{equation}

On combining (\ref{Aq1}) and (\ref{Aq22}), the definite bulk potential
field takes the form
\begin{equation}\label{Aq23}
U_{\rm b}(\rb)=U_{\rm Cd}(\rb)+\mPhi_{\rm top} .
\end{equation}
Following Bethe \cite{Beth28}, one can see that the value of $U_{\rm
Cd}(\rb)$ averaged over a unit cell cancels the last term on the
right-hand side of (\ref{Aq23}), so that for the mean bulk potential we
obtain
\begin{equation}\label{Aq24}
\bar{U}_{\rm b}=0 .
\end{equation}
Hence, the bulk potential field $U_{\rm b}(\rb)$ is independent of an
optional parameter $H$ and has no uniform component, in agreement with the
result of Ewald \cite{Ewal21}.

\section{Discussion}
It is significant that in a general case of (\ref{Aq6}) the effect of
eight sublattices mentioned above on the resulting $\mPhi_{\rm top}$ just
compensates the increase of the unit-cell volume in each of them, so that
relation (\ref{Aq22}) is reproduced with the parameters attributed to the
initial unit cell. It is also clear that the auxiliary fictitious charges
introduced in (\ref{Aq6}) vanish under periodic boundary conditions, so
that real structural charges are of importance altogether. On the other
hand, according to the above procedure of averaging, the issue (\ref{Aq7})
turns out to be indifferent to the particular definition of a unit cell.
As discussed in \cite{Khol01}, the foregoing result may also be associated
with the translational invariance as an integral property of the direct
lattice sum (\ref{Aq1}) within a special mode of summation. This
circumstance was stressed by Ewald \cite{Ewal21} as desirable upon the
definition of lattice sums as such.

Note that the potential $U_{\rm Cd}(\rb)$ as a function of $\rb$ is 
asymmetric if $H\neq0$. This is especially prominent in diatomic structures 
composed of point charges \cite{Evje32,Fise92}, but relation (\ref{Aq23}) 
retrieves the symmetric result there.

For completeness, one can show that, in terms of (\ref{Aq23}), the bulk
Coulomb energy per unit cell takes the form
\begin{equation}\label{Aq25}
\Ec_{\rm b}=\frac{1}{2}\int_{V^{\rm ini}}\rho^{\rm ini}(\rb)U_{\rm b}(\rb)
\:d\rb ,
\end{equation}
which is invariant, though $\rho^{\rm ini}(\rb)$ occupying a volume 
$V^{\rm ini}$ remains optional. According to \cite{Khol01}, the
variational derivative of $\Ec_{\rm b}$ with respect to $\rho^{\rm
ini}(\rb)$ is then equal to
\begin{equation}\label{Aq26}
\frac{\delta\Ec_{\rm b}}{\delta\rho^{\rm ini}(\rb)}=U_{\rm b}(\rb)
\end{equation}
that is the basic statement for determining $\rho^{\rm ini}(\rb)$ in a
self-consistent manner \cite{Gene87}.

It is worth noting that relations (\ref{Aq22}) and (\ref{Aq23}) also
describe the potential effect exerted by extended charges even if we deal
with spherical electronic distributions in ions \cite{Khol01}. As a
result, the potential asymmetry beyond the potential contribution of a
point-charge lattice takes place as well. According to (\ref{Aq25}), this
fact results in the asymmetry of the concentration of vacancies of
different ionic species \cite{Khol02} and so explains the $n$-type
conductivity in intrinsic semiconductors such as ZnO or GaAs
\cite{Goo269,Oate95}.

Note that the description based on (\ref{Aq1})--(\ref{Aq7}) and
(\ref{Aq22})--(\ref{Aq26}) is quite general, with including the effect
of the Lorentz field for polar unit cells as a particular case
\cite{Kho201}. A subtle problem associated with the definition of a local
polarization in ferroelectrics \cite{Vand93,Orti94} can also be
elucidated therefrom, as will be discussed elsewhere.
\vspace{-1ex}

\section{Conclusion}
\vspace{-1ex}
Without loss of generality, the problem of summation of Coulomb potentials
over crystal lattices is investigated in terms of absolutely convergent
sums with an arbitrary choice of the charge distribution in a unit cell.
The principal case of triclinic symmetry is considered. It is shown that
periodic boundary conditions imposed in an invariant manner so as to
exclude the influence of the particular choice of a unit cell are
sufficient for determining the electrostatic potentials in the bulk as
uniquely defined, with zero mean bulk potential value. A few direct
consequences of the results obtained are pointed out.
\vspace{-1ex}

\section*{Acknowledgements}
\vspace{-1ex}
I am grateful to Professor V. L. Ginzburg, Professor E. G. Maksimov and
Professor V. G. Vaks for their encouragement of this work.


\vspace{-4ex}

\begin{references}\vspace{-6ex}
\bibitem[*]{EVKhol} E-mail: kholopov@casper.che.nsk.su
\bibitem{Glas80} M.~L.~Glasser and I.~J.~Zucker, in: Theoretical
                 Chemistry: Advances and Perspectives, edited by
                 H.~Eyring and D. Henderson (Academic Press, New York,
                 1980), Vol.~5, pp.~67--139.
\bibitem{Tosi64} M.~P.~Tosi, in: Solid State Physics, edited by
                 F. Seitz and D. Turnbull (Academic Press, New York, 1964)
                 Vol.~16, pp. 1--120.
\bibitem{Harr75} F.~E.~Harris, in: Theoretical Chemistry: Advances
                 and Perspectives, edited by H.~Eyring and D.~Henderson
                 (Academic Press, New York, 1975), Vol.~1, pp.~147--218.
\bibitem{Ihm988} J.~Ihm, Rep. Prog. Phys. 51 (1988) 105.
\bibitem{Khol01} E. V. Kholopov, Preprint 2001-01 (Inst. of Inorg. Chem.,
                 Novosibirsk, 2001) (submitted to Philos. Mag. B).
\bibitem{Jack62} J.~D.~Jackson, Classical Electrodynamics (John
                 Wiley and Sons, New York, 1962).
\bibitem{Born12} M. Born and Th. von K{\'a}rm{\'a}n, Phys. Z. 8 (1912) 297.
\bibitem{Smit81} E.~R.~Smith, Proc. R. Soc. London A 375 (1981) 475.
\bibitem{Deem90} M.~W.~Deem, J.~M.~Newsam and S.~K.~Sinha, J. Phys.
                 Chem. 94 (1990) 8356.
\bibitem{Dahl65} J.~P.~Dahl, J. Phys. Chem. Solids 26 (1965) 33.
\bibitem{Redl75} A.~Redlack and J.~Grindlay, J. Phys. Chem. Solids
                 36 (1975) 73.
\bibitem{Zhan94} X.-G. Zhang, W. H. Butler, J. M. MacLaren and J. van Ek,
                 Phys. Rev. B 49 (1994) 13383.
\bibitem{Ewal21} P.~P.~Ewald, Ann. Phys. F4 64 (1921) 253.
\bibitem{Klei81} L.~Kleinman, Phys. Rev. B 24 (1981) 7412.
\bibitem{Born54} M.~Born and K.~Huang, Dynamical Theory of Crystal
                 Lattices (Clarendon Press, Oxford, 1954).
\newpage
\bibitem{Grif68} R. B. Griffiths, Phys. Rev. 176 (1968) 655.
\bibitem{Spac81} M.~A.~Spackman and R.~F.~Stewart, in: Chemical
                 Applications of Atomic and Molecular Electrostatic
                 Potentials, edited by P.~Politzer and D.~G.~Truhlar
                 (Plenum Press, New York, 1981), pp.~407--425.
\bibitem{Gene87} K.~A.~Van Genechten, W.~J.~Mortier and P.~Geerlings,
                 J. Chem. Phys. 86 (1987) 5063.
\bibitem{Task79} P.~W.~Tasker, J. Phys. C 12 (1979) 4977.
\bibitem{Wolf92} D.~Wolf, Phys. Rev. Lett. 68 (1992) 3315.
\bibitem{Evje32} H.~M.~Evjen, Phys. Rev. 39 (1932) 675.
\bibitem{Coog67} C.~K.~Coogan, Aust. J. Chem. 20 (1967) 2551.
\bibitem{Fran50} F.~C.~Frank, Philos. Mag. 41 (1950) 1287.
\bibitem{Lee980} W.~W.~Lee and S.-I.~Choi, J. Chem. Phys. 72 (1980)
                 6164.
\bibitem{Hey281} D.~M.~Heyes and F.~van~Swol, J. Chem. Phys. 75 (1981)
                 5051.
\bibitem{Coke83} H.~Coker, J. Phys. Chem. 87 (1983) 2512.
\bibitem{Beth28} H.~Bethe, Ann. Phys. F4 87 (1928) 55.
\bibitem{Fise92} I.~G.~Fisenko and E.~V.~Kholopov, Phys. Status Solidi B
                 173 (1992) 515.
\bibitem{Khol02} E. V. Kholopov, Zh. Struc. Chim. (to be published).
\bibitem{Goo269} W.~Van Gool and A.~G.~Piken, J. Mater. Sci. 4 (1969) 105.
\bibitem{Oate95} W.~A.~Oates, G.~Eriksson and H.~Wenzl, J. Alloys
                 Comp. 220 (1995) 48.
\bibitem{Kho201} E. V. Kholopov, Preprint 2001-02 (Inst. of Inorg. Chem.,
                 Novosibirsk, 2001) (submitted to Philos. Mag. B).
\bibitem{Vand93} D. Vanderbilt and R. D. King-Smith, Phys. Rev. B 48
                 (1993) 4442.
\bibitem{Orti94} G. Ortiz and R. M. Martin, Phys. Rev. B 49 (1994) 14202.
\end{references}
\end{document}